\documentclass[12pt,conference,a4paper, twocolumn, balance]{article}
\usepackage{times,amsmath, url, paralist}
\usepackage[doi=false, isbn=false, url=false]
{biblatex}
\bibliography{literature}
\usepackage[margin=1in]{geometry}
\usepackage{balance}	
\usepackage{setspace}

\usepackage{subfig}
\usepackage{float}
\floatstyle{plaintop}
\restylefloat{table}

\makeatletter

\newcommand*{\TitleFont}{%
      \usefont{\encodingdefault}{\rmdefault}{b}{n}%
      \fontsize{14}{20}%
      \selectfont}

\title{\vspace*{-25pt}\TitleFont\bf Wisdom of Crowds Algorithm for Stock Market Predictions}
\singlespace{
\author{
\vspace{-5pt}
\fontsize{12}{12}\selectfont 
{Marko Velic{$~^{1}$}, Toni Grzinic{\small $~^{2}$}, Ivan Padavic{\small $~^{3}$}}
\\
\vspace{-5pt}
\fontsize{12}{12}\selectfont\itshape
$^{1}$\,University Computing Centre, Josipa Marohnica 5, 10000 Zagreb, Croatia\\
\vspace{-5pt}
\fontsize{12}{12}\selectfont\itshape
$^{2}$\,Croatian Academic and Research Network, Josipa Marohnica 5, 10000 Zagreb, Croatia\\
\vspace{-5pt}
\fontsize{12}{12}\selectfont\itshape
$^{3}$\,Trikoder Ltd., Draskoviceva 80, 10000 Zagreb, Croatia \\
\fontsize{12}{12}\selectfont\itshape
E-mail(s): marko@velic.biz, toni.grzinic@carnet.hr, ivan.padavic@trikoder.net
}}
\date{}

\begin{document}

\setlength{\belowdisplayskip}{12pt} \setlength{\belowdisplayshortskip}{12pt}
\setlength{\abovedisplayskip}{12pt} \setlength{\abovedisplayshortskip}{12pt}

\maketitle
\pagestyle{empty}
\thispagestyle{empty}
\noindent\textbf{{\fontsize{12}{12}Abstract.}} \emph{In this paper we present a mathematical model for collaborative filtering implementation in stock market predictions. In popular literature collaborative filtering, also known as Wisdom of Crowds, assumes that group has a greater knowledge than the individual while each individual can improve group’s performance by its specific information input. There are commercially available tools for collaborative stock market predictions and patent protected web-based software solutions. Mathematics that lies behind those algorithms is not disclosed in the literature, so the presented model and algorithmic implementation are the main contributions of this work.}\\
\emph{ \ }\\
\emph{}\textbf{\fontsize{12}{12} Keywords.} 
Collaborative Filtering, Wisdom of Crowds, Crowdsourcing, Stock, Market, Prediction
\\
\\
\textbf{\fontsize{12}{12} 1.\ Introduction}\\
 \ 
\\
\indent In 1906, during the West of England Fat Stock and Poultry Exhibition, Francis Galton discovered the mathematical and statistical patterns of group average estimations and their advantages over individual ones. At the event, a group of estimators successfully provided an average weight value of an ox remarkably close to the real value \cite{Galton}. This phenomenon is a foundation of the group estimating systems that we call collaborative filtering (CF) systems.
\\
\indent CF systems, which are also known by the more marketable terms Wisdom of Crowds and Crowdsourcing, are implemented in various domains like systems for rating books, movies, music and stocks \cite{MotleyCAPS:2013:Online}.
\\
\indent Modern CF considerations origin from the early 90s of the 20th century \cite{Cohen:1992} when the term CF was coined by Goldberg et al \cite{Goldberg:1992}. The basic assumption is that group of individuals can yield more knowledge than an individual alone while at the same time each individual can probably improve group's performance by contributing with its specific small chunks of knowledge. \\
\indent One interesting CF application is a very popular stock rating system that is operational since 2006 and its predictive power is confirmed in literature \cite{Hill:2011}. The system is patent protected\footnote{US patent 7813986 and 7882006}. Those patents cover various features while providing limited considerations about the algorithm that lies underneath the rating system. As it is stated on the project's website, the algorithm is kept secret and according to developers it is a subject of constant improvements, upgrades and fine tuning. \\
\indent This paper presents a model designed to work in the similar fashion and to yield similar results, having in mind not to make a copy of the mentioned system. Existing system's help was used as a general guidance for developing the presented mathematical model.
\indent
\\
\\
\textbf{\fontsize{12}{12} 2.\ Predicting Stock Markets}\\
\ \\
\indent Stock Market forecasting is popular and attractive for both laymen and scientists. Besides widely used and economically confirmed fundamental analysis of the companies' performance, the challenge is to create an artificial intelligence (AI) model that will predict  trends and events. Due to that a wide range of techniques was adopted, tested and some interesting results were achieved. From technical analysis considerations, fuzzy logic \cite{Chen2007}, neural networks \cite{Schoneburg1990} to hybrid approaches \cite{Hadavandi2010}. Many of the presented models achieved positive results in a short term. Since stock market is subject of constant changes and various influences, it is a greater challenge to achieve performance in a long term. \\
\indent There is a known logical problem that implementation of the prediction model, thus interference of the agents using the model, causes degradation in performance of the model itself since system behaviour is changed in a global perspective. It is known that algorithms used in automatic trading have a lifespan measured in weeks, both because of their solely influence on the market and their influence on each other when more algorithms work on the same market. Similar problems with possible impacts on the market, and not just on the model performance, are utilizations of the High frequency Trading Systems (HFT)\footnote{International Organization of Securities Commissions (IOSCO) - Regulatory Issues Raised by the Impact of Technological Changes on Market Integrity and Efficiency}. Following these considerations we can conclude that some form of feedback and system self-correction is needed to ensure long term model performance.\\
\indent CF approach offers a solution to this problem in a form of constant monitoring and evaluation of all agents in the system and thus self-adapting the prediction system. This type of predictive model will be described in the following chapters. 
\\ \\
 \textbf{\fontsize{12}{12} 3.\ Collaborative Filtering Approaches}\\
 \ \\
 \indent CF systems are divided into two main categories, memory-based (or user based) and model-based (item-based) \cite{Sarwar:2001}.
Former utilize the entire user-item database for prediction. These systems find neighbours between clusters of user-item pairs. On the other hand model based systems develop a model for recommendation based on history of user ratings. Common problem in item based systems is a "cold start" issue where users must rate sufficient items to induce the system to make right predictions \cite{Schein:2002}. \\
\indent Group lens \cite{Herlocker:1999} suggested an automated CF model for personalized usenet messages and Pearson correlation was used to weight user similarity. An extension of the Group lens method was the use of Spearman rank correlation coefficient instead of Pearson correlation. Spearman correlation does not rely on model assumptions but correlates ranks instead of rating values. These memory based algorithms are the most widely implemented \cite{McLaughlin:2004}. \\
\indent Popular commercial CF systems are Amazon \cite{Linden2003} and Netflix.  Amazon is an e-commerce company that introduced item-to-item based filtering, a proprietary algorithm that generates recommendations based on the similarity of items\footnote{US patent 6,226,649}. Amazon's approach was the reaction to, at that time, inefficient user based systems. Netflix is an on-demand streaming media which maintains a personalized video-recommendation system based on users' ratings and reviews. In 2006 the company launched a competition with a bounty of 1M dollars for a 10\% improvement of their CF algorithm.
\\
\indent Some psychological centric works  suggest that it is in human nature to estimate the boundaries of predictions \cite{griffiths2006}, \cite{Mozer2008}. Griffins et. al conducted tests on everyday problems like estimating human life spans, where they found that individuals make very similar predictions to probabilistic models \cite{griffiths2006}. The most known pitfall in group relations is the groupthink phenomenon, where the group tries to minimize conflict making consensus decisions \cite{Janis1982}. \\
\indent Groupthink also relates to user relations in CF systems because users' votes, reviews and thoughts can influence other users.
\\
\textbf{\fontsize{12}{12}4.\ Algorithm}
 
 \ \\ 
 \textbf{\fontsize{12}{12} 4.1.\ Rating system}\\
 \ \\
\indent We define a set of stocks used in game
$St = \{ st_1, st_2, \dots, st_n \}$
where $n$ is a number of stocks in a game.
Stock price i.e. value is considered through time where $ t_0 $ is start time of interest, e.g. prediction insertion, and $ t_i $ is observed moment so 
$ t_i = t_0 + \Delta t $ where $ \Delta t $ represents time passed.
We define stock value at the starting time and at the observed time as $ V_{S_{t_0}} $ and $ V_{S_{t_1}} $ respectively. System gets its inputs from a set of players $Pl$, where each player $pl$ enters predictions.
Stock gain is defined as
\begin{equation}
\Delta St_{i} = (\frac{V_{St_{i}}}{V_{St_{0}}}) * 100
\end{equation}
\indent In the same manner, we define index values as $ V_{It_0} $ and $ V_{It_1} $ for starting value and value at the observed moment and index gain is defined as 
\begin{equation}
\Delta It_{i} = (\frac{V_{It_{i}}}{V_{It_{0}}}) * 100
\end{equation}
\indent Each pick has an orientation $o$ from a set $O$ that can represent outperformance or underperformance of the stock compared to the selected index $ O = \{1,2\} $ respectively.

\indent To achieve a one-line calculation for a prediction score we introduce two multipliers:
\begin{equation}
M_{\Delta S}=
\begin{cases}
1, &o = 1;\\
-1, &o \neq 1
\end{cases}
\end{equation}
\begin{equation}
M_{\Delta I}=
\begin{cases}
1, &o = 1;\\
-1, &o \neq 1
\end{cases}
\end{equation}

\indent Score for the individual prediction is calculated as
\begin{equation}
\varsigma_{pr}= \Delta St_{i} * M_{\Delta S} + \Delta It_{i}  * M_{\Delta I}
\end{equation}

We define player's score as
\begin{equation}
\varsigma_{pr} = \Sigma \varsigma_{pr|pl}, t> t_{0+\Delta P}
\end{equation}
where $t_{0+\Delta P}$ connotes moment of the pending period end for a particular stock prediction. Pending period is introduced to eliminate possible short-term manipulations.

\indent To calculate accuracy for a given player we define a set of player's predictions $Pr=\{pr_1,pr_2,\dots, pr_n\}$. Each player has its own predictions $Pr_i$. Global predictions count of a given player is $\vert Pr \vert $ and positive predictions count $\vert Pr^+ \vert $ is number of predictions in which player achieved positive score ${Pr}^+ \subseteq Pr$.
\begin{equation}
{Pr}^+=\{pr  :  pr \in \{pr : \varsigma_{pr} > 0 \}\}
\end{equation}
\indent In order to minimize computational requirements, we normalize each player's score to 100.
To find player's prediction accuracy we need average number of predictions for all players $N_{pl}$ :
\begin{equation}
N_{pl}=\dfrac{\vert Pr \vert}{\vert Pl \vert}
\end{equation}

\indent To calculate accuracies of all active players $A = \{\alpha_1, \alpha_2, \dots, \alpha_n\}$ we need mean accuracy $\bar{A}$ for all players
\begin{equation}
\bar{A}=\dfrac{\sum_{i=1}^{n} \alpha_i}{N_{pl}}
\end{equation}

\indent Furthermore, we define normalized prediction count for each player as 
\begin{equation}
\vert \widehat{Pr}_i \vert = 
\begin{cases}
\vert Pr_i \vert, &\vert Pr_i \vert \leq 100;\\
100, &\vert Pr_i \vert > 100.
\end{cases}
\end{equation}

and normalized positive pick count
\begin{equation}
\vert \widehat{Pr}^+ \vert = 
\begin{cases}
\vert \widehat{Pr}^+ \vert, &\vert Pr_i \vert \leq 100;\\
\vert \widehat{Pr}_i \vert,  &\vert Pr_i \vert > 100. 
\end{cases}
\end{equation}

\indent We have normalized positive predictions count for a player as
\begin{equation}
{\widehat{Pr}}^+=\dfrac{100}{\vert Pr \vert} * \vert {Pr}^+ \vert
\end{equation}

\indent Individual normalized accuracy for player $i$ is defined
\begin{equation}
\alpha_i=\dfrac{\vert \widehat{Pr}^+ \vert}{\vert Pr \vert}*100
\end{equation}

\indent When aligned with other players' accuracies we get individual rating
\begin{equation}
R_T=\vert Pr \vert * \alpha_i
\end{equation}
\indent Using Bayesian average we can calculate accuracy rank i.e. probability of player's accuracy for those with less than 100 picks as
\begin{equation}
R_A=\dfrac{\bar{\vert Pr \vert}*\bar{A} + R_T}{\vert Pr \vert + \bar{\vert Pr \vert}}
\end{equation}

\indent To rank all players we consider their scores and accuracies jointly. First we calculate player's score rating as a percentile rank for player $i$. After sorting ascendantly, score and accuracy percentile ranks are defined:
\begin{eqnarray}
y_{Si}=\dfrac{pl_i}{\vert \varsigma_{pl} \vert} * 100
\\
y_{Ai}=\dfrac{pl_i}{\vert A \vert} * 100
\end{eqnarray}
\indent Raw rating for each player is calculated as:
\begin{equation}
r_i = 2/3 * y_{Si} + 1/3 * y_{Ai}
\end{equation}
and we have a sorted set of ranks $R = \{r_1, r_2, \dots, r_n\} : r_i < r_{i+1}$ for each player $i$.
Due to analogy raw ranks percentiles are defined
\begin{equation}
y_{Ri}=\dfrac{pl_i}{\vert R \vert} * 100.
\end{equation}

\indent When we have calculated players' ratings, the next task is to calculate stock ratings accordingly. To ensure fraud resistance, we implement following minimal number of predictions $_{min}\vert Pr_{St_{i}} \vert = 5$, and at least one player must be among top 40\% of all players $_{min} Y_{Ri} = 60$. Stocks that satisfy minimal conditions are denoted $st_i'$ where $St' \subseteq St$. Therefore, we find players that made predictions for the given stock 
$Pl' \subseteq Pl$.
Then we apply the qualification algorithm
\begin{eqnarray}
\exists St_i' \Longrightarrow (\vert Pr_{St_{i}} \vert > _{min}\vert Pr_{St_{i}}\vert) \wedge \nonumber \\ (\exists p_{Li} \Longrightarrow (y_{Ri} > _{min} Y_{Ri} = 60)).
\end{eqnarray}

\indent Every qualified stock has at least one positive or one negative prediction among all its predictions $\forall st_i' \Longrightarrow Y_{RO} \vee Y_{RU}$. $Y_{RO}$ is defined as outperform prediction, while $Y_{RU}$ are underperformed predictions. Taking into account players rating to rate the particular stocks we use the sum of stock predictions
\begin{equation}
\label{eq_ratings}
\sum_{RATINGS}= \sum_{i}^{\vert RO \vert} y_{ROi} + \sum_{i}^{\vert RU \vert} y_{RUi}
\end{equation}

thus stock score $ \theta_i $ is
\begin{equation}
\theta_i = \frac{\sum_{i}^{\vert RO \vert} y_{ROi}}{\sum _{RATINGS}}
\end{equation}

\indent We calculate percentile ranks ordering stock scores $\Theta = \{\theta1, \theta2, \dots , \theta_p \}: \theta_i < \theta_{i+1}$
and percentile rank of a particular stock 
\begin{equation}
y_{\Theta i} = \frac{i}{\vert \Theta \vert} * 100
\end{equation}
 
  \ \\
 \textbf{\fontsize{12}{12} 4.2.\ Keeping it fair}\\
 \ \\ 
\indent In order to prevent manipulation, besides delayed stock prices data, we have implemented several limitations to keep the prediction system fair. Limits include prediction maturity i.e. time span after a prediction starts affecting player's score, minimal number of active and mature predictions for a stock to have a rating, minimal rank for at least one player that entered his prediction for a stock to have a rating, number of active and mature predictions for a player to have a rank and finally minimal score that prediction with positive score must achieve to start counting for a player's overall score.
\indent
\\
\\
\\
\\
\\
\textbf{\fontsize{12}{12} 5.\ Evaluation}\\
\ \\
\indent As it is mentioned earlier, one of the problems of CF systems is a cold start phenomenon. Same problem exists with our implementation. At the moment of writing of this paper, system is operational for more than six months with 47 active players. Total number of entered predictions was 667, half of them were active for 130 stocks. 
\\
\indent Since both the elapsed period and players count are rather small, due to the system's cold start, no reliable statistical evaluation could be made. Here we give current gains of the top 10 stocks identified by the system just for illustration. Results are presented in Table \ref{top10}. One possible criticism of the presented algorithm can be a possible occurrence of the groupthink phenomenon. Players that are not so experienced could imitate high ranked players and predict in the same way. The main concern is if the time component solely is enough insurance, since time-delayed imitating of the high ranked player can not yield the same result i.e. imitator will always stay one step behind. Also, if a high ranked player makes a mistake, negative score is cascading on the imitators too.
\\

\begin{table}[H]
	\centering
	\caption{\bf Top 10 stocks by the system's rating compared to their yearly and monthly gain}
    \begin{tabular}{|l||l|l|l|}
        \hline
        Ticker   & Rank  & 1 Y(\%)     & 1 M(\%)   \\ \hline
        VDKT-R-A & 95.83 & 238.83 & 37.20 \\ 
        TISK-R-A & 94.44 & 66.47  & 16.04 \\ 
        PBZ-R-A  & 93.06 & 10.22  & 3.00  \\ 
        LPLH-R-A & 91.67 & -51.68 & -3.90 \\ 
        KODT-R-A & 90.28 & 2.89   & 16.92 \\ 
        ISTT-R-A & 88.89 & -20.84 & -6.58 \\ 
        HUPZ-R-A & 87.50 & 16.83  & 11.88 \\ 
        CROS-P-A & 86.11 & 66.24  & 3.17  \\ 
        VART-R-1 & 80.56 & -32.08 & 20.62 \\
        SLRS-R-A & 79.17 & 78.53 & 17.37 \\
        \hline
    \end{tabular}
	\label{top10}
\end{table} 

\indent
\\
\\
\\
\textbf{\fontsize{12}{12} 6.\ Future Work}
 \\ \\
 \textbf{\fontsize{12}{12} 6.1.\ Testing}\\
 \\
\indent Besides fine-tuning of the proposed algorithm, future work will cover measurement of the algorithm’s performance, identification of the groupthink problem and mitigation.
 \\ \\ 
 \textbf{\fontsize{12}{12} 6.2.\ Fine Tuning}\\
 \\
\indent Based on the testing results, further improvements can be made on the algorithm. One of the most obvious fine tuning opportunities is an improvement of the final stock rating algorithm where individual player ratings are considered when calculating the stock score. As can be seen in formula \eqref{eq_ratings} all players' ratings are taken as they are. Here a simple exponential or similar transformation can be used to increase the weight for higher ranked players.
 \indent
 
    \ \\
 \textbf{\fontsize{12}{12} 6.3.\ Algorithmic Trading Considerations}\\
 \ \\
\indent Since the presented system implements feedback i.e. constant monitoring of each player's performance and correction of its influence on the stock rating, question is if it can be used to achieve positive trading results of an AI trader. Outperforming predictions from highly ranked players could be used as an indicators for buying and under performing predictions for short-selling. In that way positive results could hypothetically be achieved from both directions.
\indent 
\\
\\
\textbf{\fontsize{12}{12} 7.\ Conclusion}\\
\ \\
\indent There is a probative evidence that CF can generate valuable knowledge from the groups of people about the specific topic. That kind of knowledge extraction mechanism found its use in various domains like entertainment where users can rate movies or music and more serious domains like education and business where users rate books or stocks. \\ \indent Stock rating systems exist on the market and their software implementations are described and protected by patents but mathematics and algorithms underneath are not publicly available. This paper describes one possible way of implementing CF idea for stock market predictions. The described implementation is very recent so quality assessment of the predictive model will be a subject of future research. 
\\ \\
\textbf{\fontsize{12}{12} 8.\ Acknowledgement}\\
\ \\
Presented algorithm is a part of the iCapital project \cite{Icapital:Online} that is property of the InterCapital Inc. The algorithm is presented with permission.

\balance

\renewcommand{\refname	}{\@setfontsize\normalsize{12}{12} References}
\printbibliography

\end{document}